\documentclass[prl,reprint,noeprint,amsmath,amssymb,aps,longbibliography]{revtex4-2}
\usepackage{amsmath}
\usepackage{array}[=2016-10-06]
\usepackage{graphicx}
\usepackage{xcolor}
\usepackage{upgreek}
\usepackage{dcolumn}
\usepackage{bm}
\usepackage{color}
\usepackage{mathtools}
\usepackage[normalem]{ulem}
\usepackage[english]{babel}
\usepackage{xr-hyper}
\usepackage[utf8]{inputenc}
\usepackage[T1]{fontenc}
\usepackage[colorlinks=true,linkcolor=blue,urlcolor=blue,citecolor=blue]{hyperref}

\newcommand{\fig}[1]{Fig.~\ref{#1}}

\newcommand{\eq}[1]{Eq.~(\ref{#1})}

\newcommand{\squarebk}[1]{\left[{#1}\right]}

\newcommand{\anglebk}[1]{\left\langle{#1}\right\rangle}

\newcommand{\dE}{\Delta E}

\begin{document}
\title{Two-stage Slowdown of Vacancy-mediated Diffusion \\in a Lattice Gas Model with Quenched Disorder and Spins}

\author{Qiang Zhai} 
\email{qiangzhai303@163.com}

\address{
Department of Physics, Yunnan Normal University, Kunming, Yunnan, 650500, China\\
}

\date{\today}

\begin{abstract}
We present a lattice gas model with random 
interactions quenched in space and spins quenched to the particles.
The vacancy-mediated diffusion is studied in equilibrium. A two-stage
slowdown is witnessed in both the mean squared displacement
$\langle \Delta \bm{r}_i^2(t) \rangle$ and the spatial overlap function $q(t)$.
We demonstrate that the plateaus are due to anti-correlation at the corresponding length scale. The anti-correlation
diminishes quickly with increasing lag time.
A broad and yet finite distribution of the survival function of vacancies suggests their normal diffusion at the long-time limit.
Our model can serve as a new playground for understanding hierarchical dynamics in glass-like dynamical systems.

\end{abstract}

\maketitle

\emph{Introduction.---}Relaxation and diffusion are of central interest to the study of glassy
systems~\cite{Parisi_2010,Berthier_2011,Charbonneau_2017,Dahlberg_2025}, crowded
biological media~\cite{Wong_2004,Bronstein_2009,H_fling_2013} and other heterogeneous
environment~\cite{Bouchaud_1990}. In the context of equilibrium dynamics, relaxation characterizes the memory of the
system and diffusion measures
the spatial spread of individual components.  Sluggish relaxations have been reported for overlap functions in
spin glasses and supercooled liquids. The diffusion can be categorized by the exponent $\alpha$ of the asymptotic
time dependence of mean-squared displacement (MSD) $\langle \bm{r}^2(t) \rangle \sim t^\alpha$. The diffusion is
subdiffusive, normal or superdiffusive given $\alpha<1$, $\alpha=1$ or $\alpha>1$ respectively. Both relaxation and
diffusion reveal important aspects of the dynamics, e.g. caging~\cite{Weeks_2002},
memory~\cite{Scalliet_2019,Paga_2024}, and the level of
crowdedness~\cite{Weiss_2004}.

Lattice models~\cite{Fredrickson_1985,Nicodemi_1997a,Nicodemi_1997b,Arenzon_2000,Fierro_2000,de_Candia_2001,Denny_2003,Ritort_2003,Bertin_2005,Crisanti_2011,Tondl_2014,Zhang_2017,Nishikawa_2024}
are valuable tools in understanding the dynamics in complex media.
Among them are vacancy-mediated diffusion
models, which have been intensively studied \cite{Glarum_1960,Phillips_1972,Bordewijk_1975,Tondl_2014,Zhang_2017} partially because of
their intuitive physical interpretations. Free-volume hole or defect creates soft spot facilitating subsequent molecular
motions. Here we present a vacancy-mediated lattice gas model and examine its dynamics. We find a two-stage slowdown of the relaxation and the diffusion manifested as plateau in the spatial overlap function and MSD. Through the
lens of the motion of vacancies in real space, we show spatial confinement at different length scale. This hierarchical
dynamics is further corroborated by strong anti-correlations at the corresponding length scales, where the plateau
emerges. Our findings are clear demonstration of the hierarchical dynamics in glassy systems with separation of length
scales.

\emph{The lattice model and methods.}---
The lattice model is defined on a squared lattice of 2D of linear size $L$. The energy of the model is given by,
\begin{equation}\label{eq:E}
E=\sum_{<i,j>}J_{ij}n_in_js_{p_i}s_{p_j}~,
\end{equation}
under the constraint $\sum_{i}n_i=L^2\rho$, where $\rho$ is the particle density.
The interaction $J_{ij} = \pm 1$ is drawn randomly from a bimodal distributions with equal probability and spatially quenched for bonding
between lattice site $i$ and $j$ of nearest neighbors. Occupation of a lattice site either by a particle or a vacancy (P/V) is indicated by $n_i$. If the lattice site is occupied by a particle $n_i=1$, otherwise $n_i=0$. The spin variables
are sampled randomly from a bimodal distribution with $s_{p_i}=\pm 1$. Notice that the subindex $p_i$ is the label of
particle at lattice site $i$.
In order to track the trajectories of the P/V, all the P/V are labeled uniquely.   \emph{Importantly, the spin variables are time-invariant}, i.e., quenched to the particles at the start of the simulation. Particle hopping is allowed when at least one of its nearest neighbor is occupied by a vacancy. The diffusion dynamics then generate a new particle configuration and update the energy in \eq{eq:E}. 
We have used $L=128$ in our simulations. To study glassy dynamics in a crowded environment, the particle density $\rho$ is set to 0.99.

We have adopted a mixture of parallel
tempering~\cite{Hukushima_1996} and
non-local swap algorithm~\cite{Grigera_2001} to equilibrate the system down to $T=0.22$. The parallel tempering technique is applied to replicas 
(samples with fixed $J_{ij}$ and $s_{p_i}$ but varying initial particle configurations). During this stage, nonlocal swapping among 
all particles and vacancies are performed for fast equilibration. After the system are equilibrated, we simulate the vacancy-mediated diffusion with Metropolis algorithm.
The particles can hop to a vacant site at the rate,
\begin{equation}\label{eq:dE}
w=w_{0} \exp {\squarebk{-\dE \Theta(\dE)/k_BT}}
\end{equation}
where $\Theta(x)$ is the usual Heaviside function. The Boltzmann constant $k_B$ is set to unity for simplicity.
The hopping attempt frequency $w_{0}=1\times10^6$. The energy difference $\dE$ is determined by the spin configurations
before and after the particle hopping and can be evaluated by \eq{eq:E}. We study the dynamics in the temperature range
$0.22 \le T \le 1.10$. To enhance the computational efficiency, the rejection-free approach is used~\cite{Bortz_1975}. 
Detailed balance is guaranteed by the algorithms used in our simulations.

\begin{figure}[!th]
\includegraphics[width=0.8\columnwidth]{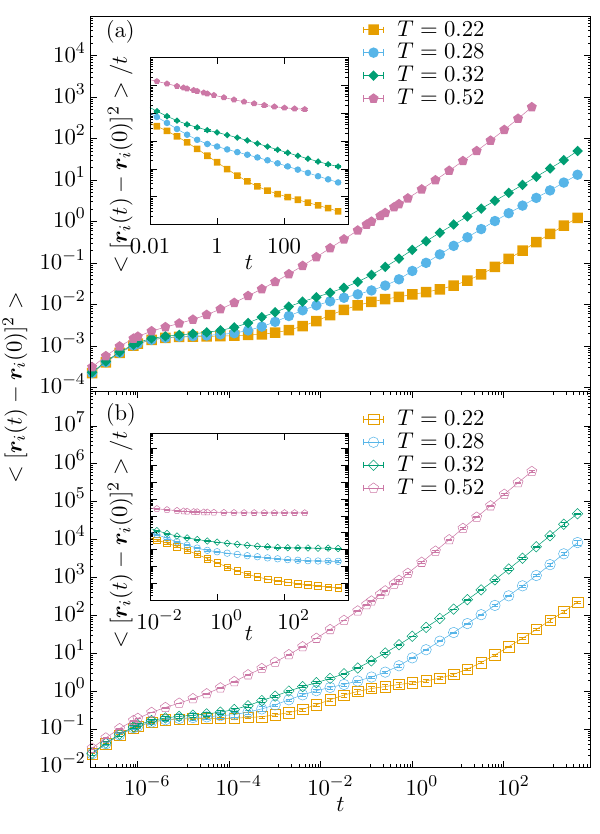}
\caption{The ensemble averaged mean-squared displacement for particles (a) and vacancies (b).A two-stage slowdown is seen for both particles (filled symbols) and vacancies (hollow symbols). The inset of (a) and (b) displays the ratio of MSD and time. The lines are smooth polynomial fit, serving as guides to the eyes.}\label{fig:msd}
\end{figure}

\emph{A two-stage slowdown of relaxation and diffusion .---} The ensemble-averaged MSD  is defined as, $\overline{\anglebk{[\bm{r}_i(t)-\bm{r}_i(0)]^2}}$, where the brackets indicate thermal average and the bar indicates disorder average. \fig{fig:msd} displays MSD for both particles (a) and vacancies (b). Although the magnitude of MSD differs between particles and vacancies, they share similar profiles for the temporal evolution. The most interesting features emerge at low temperatures. For example, at $T=0.22$, the slowdown is seen. The first plateau starts around $t \approx 2 \times 10^{-6}$ and ends near $t \approx 2 \times 10^{-4}$ and then the second one develops around  $t \approx 2 \times 10^{-2}$. The precursor of the second slowdown can also be found at higher $T$. Note that the second slowdown never develops into a true plateau even for the lowest temperature we have studied. The first plateau is commonly observed in glassy systems while the second slowdown has only been reported in a few studies~\cite{Crisanti_2011,Shiraishi_2023}.

Another interesting feature of the MSD is that it demonstrates a rich combination of diffusion regimes. Diffusion processes are categorized into different classes based on the asymptotic scaling behavior of $\mathrm{MSD} \sim t^\alpha$. Subdiffusion is distinguished by $\alpha <1$ and normal diffusion is characterized by $\alpha  =1$. For $T=0.52$, only the first slowdown is observed and the diffusion turns normal after a long period of sub-diffusive regime. This is found for both particles and vacancies. As $T$ is lowered, vacancies can diffuse normally down to $T=0.32$ but particles remains subdiffusive for the longest time that we have measured. The inset of \fig{fig:msd} shows the ratio of MSD and $t$ at long time. The flatness of the ratio indicates normal diffusion.      

\begin{figure}[!th]
\includegraphics[width=0.8\columnwidth]{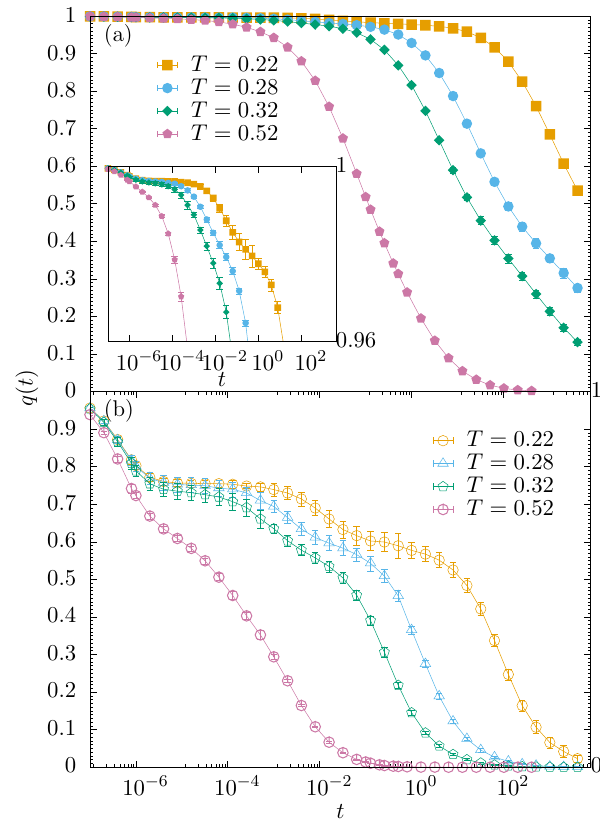}
\caption{The spatial overlap function $q(t)$ for particles (a) and vacancies (b). Inset of (a) is the magnified portion of $q(t)$ at the early stage of the relaxation.}\label{fig:sisf}
\end{figure}

\begin{figure*}[!th]
\includegraphics[width=0.66\columnwidth]{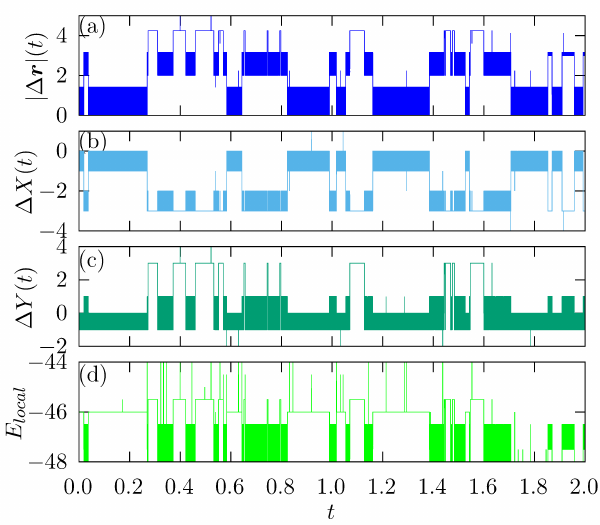}
\includegraphics[width=0.66\columnwidth]{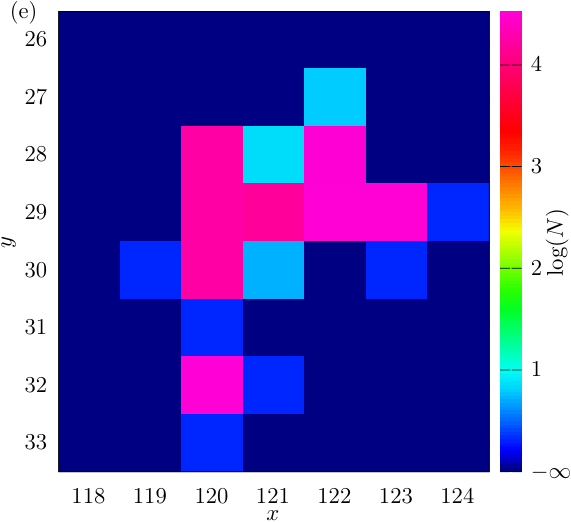}
\includegraphics[width=0.66\columnwidth]{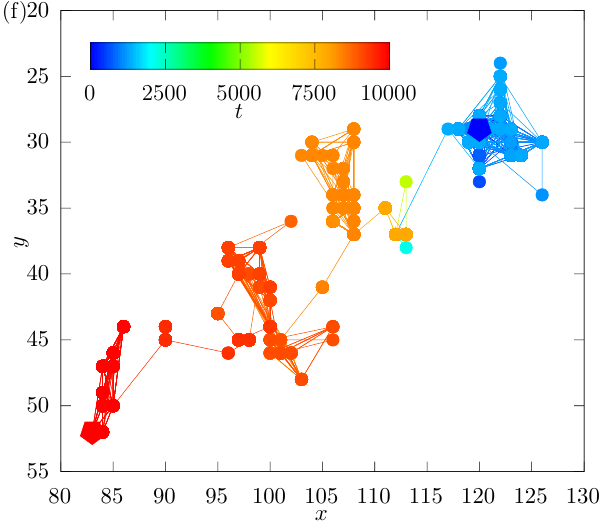}
\caption{The typical motion of a vacancy at $T=0.22$, with $\Delta \bm{r} (t)$ in (a),  $\Delta {X}(t)$ in (b) and $\Delta {Y}(t) $ in (c). The visiting numbers $N$ of this vacancy to each lattice site from $t=0$ to $t=2$ are visualized in (e),  the energy temporal evolution of the local region displayed in (e) is given in (d). Trajectory of this vacancy is plotted up to $t=10000$ in (f).}\label{fig:real}
\end{figure*}

To better assess the two-stage slowdown, we also study the relaxation dynamics. The relaxation process is quantified by the spatial overlap function of the from,
\begin{equation}
q(t)=\overline{\anglebk{1-\Theta[|\bm{r}_i(t)-\bm{r}_i(0)|]}}~,
\end{equation}
where $\Theta(x)=1$ if $x>0$. Thus, $q(t)$ measures the probability that a particle has zero net displacement for the duration $t$ and has been shown to serve the similar purpose of self-intermediate-scattering function for understanding structural relaxation~\cite{Lulli_2020}. \fig{fig:sisf} shows $q(t)$ for both particles and vacancies. The relaxation curves demonstrate similar multi-stage profile as already observed in MSD. One also  needs to carefully magnify the relaxation curve of particles to be able to see the details of the two-stage slowdown (inset of \fig{fig:sisf} (a)). Importantly, though measured with different quantity, all the slowdown occurs at the same time for both MSD and $q(t)$, implying that the underlying mechanism must be the same.

\emph{A tour to the real space.---} In order to understand the dynamical slowdown, let us focus on the motion of a typical vacancy first and examine the two-stage slowdown in real space. \fig{fig:real} exhibits the motion of a vacancy at $T=0.22$ from different aspects. From \fig{fig:real} (a) to (c), $|\Delta \bm{r} (t)|$ and its two components $\Delta X(t)$ and $\Delta Y(t)$ are shown up to $t=2$.  Particle localization at different spatial and temporal scales is readily seen and can be classified into two types. Type I: for short timescale of order $1 \times 10^{-6}$, the vacancies are trapped in a relative-small region, bouncing back and forth. Type II: for the timescale of unity, the vacancy can escape the localization of the first type, but get trapped again before traveling between these two traps, i.e., smaller traps within one larger trap. We believe that the persistent motions of Type I gives rise to the first plateau. Transitions between small traps break the synchronization, leading to the rise of MSD from the first plateau. Motions of Type II are responsible for the second  slowdown observed in both MSD and $q(t)$. This argument will be further quantified in the next section.  Further, we record the numbers $N$ that a site has been visited by the vacancy and plot them in \fig{fig:real} (e). Two red-ish groups of the lattice sites correspond to the Type I localization. The local energy evolution $E_\mathrm{local}$ of the plotted region in \fig{fig:real} (e) is evaluated and displayed in \fig{fig:real} (d),  two preferred energy levels are observed. The rare spikes in $E_\mathrm{local}$ are due to the visits of neighboring vacancies. These energy levels might be local free energy minimum attracting the vacancy. \fig{fig:real} (f) plots the coarse-grained trajectory of the same vacancy with color indicating the time. The locally-explored regions are dense and connected by rare jumps. Notice that from $t=0$ to $t=2500$, the vacancy is trapped in one local region (Type II), which consists of the two small traps (Type I) shown in \fig{fig:real} (e). In contrast, after around $t=5000$, another three different regions (Type II) are visited by the same vacancy. This indicates that a broad distribution of waiting time must exist for such hopping process and that the rare jumps connecting the the Type II clusters are responsible for the end of the second dynamical slowdown.

\begin{figure}[!th]
\includegraphics[width=\columnwidth]{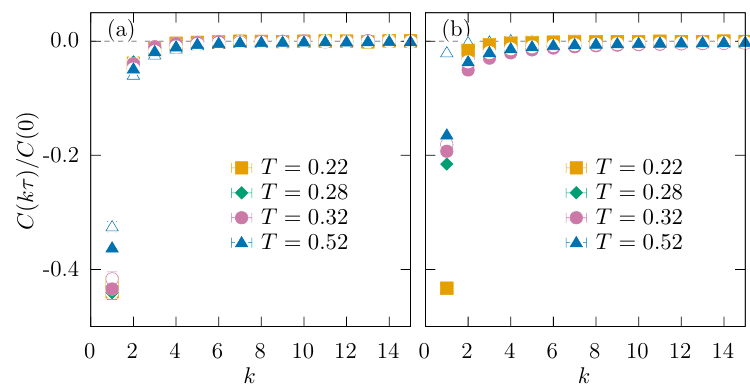}
\caption{The displacement correlation $C(k\tau)/C(0)$  defined by \eq{eq:anti} for (a) $\tau =2\times 10^{-6}$ and (b) $\tau =1$. The solid symbols indicate the anti-correlation of the particles, and the hollow symbols are for the vacancies.}\label{fig:anti}
\end{figure}
\emph{Quantifying the slowdown.---} We now quantify the correlation of hopping motion for P/V.  The displacement correlation function of fixed time-interval is defined by,
\begin{equation} \label{eq:anti}
\begin{aligned}
C(k\tau)&=\overline {\langle \delta \bm{r}_i(0)\delta \bm{r}_i(k\tau) \rangle},\\
\delta \bm{r}_i(k \tau) &= \bm{r}_i[(k+1)\tau]-\bm{r}_i(k\tau)
\end{aligned}
\end{equation}
where $k$ is the lag and $\tau$ is the fixed-time interval.  \fig{fig:anti} (a) and (b) plot the normalized $C(k\tau)$ at $\tau=2\times 10^{-6}$ and $\tau=1$. The two time intervals are chosen to approximately match the timescale of the two slowdown regime of MSD and $q(t)$ at $T=0.22$. For $\tau=2 \times 10^{-6}$,  strong anti-correlations are shown for both particles and vacancies. Anti-correlation measures the tendency of P/V to hop backwards. From $T=0.22$ to $T=0.32$, the magnitude of anti-correlation overlaps, larger than that of $T=0.52$.   This is consistent with the first-plateau shown in \fig{fig:msd} and \fig{fig:sisf}, where  a stronger slowdown is seen for the former three temperatures.  In  contrast, for $\tau=1$, the anti-correlation remains strong at $T=0.22$ and weakens for the other temperatures. Indeed, only at $T=0.22$ the second slowdown is visually evident.  The strong anti-correlation at $T=0.22$ for both $\tau$ echos the picture in real space, i.e. being trapped at two different length scale.  

\begin{table}[!htb]
	\caption{The maximum waiting time $\tau_\mathrm{w}$ and total measurement time $t_\mathrm{ob}$ for P/V of the hopping-event-based simulation. For particles, the listed maximum values are for mobile particles, the values in the parenthesis equal the measurement time, indicating that there are immobile particles during the entire measuring time window.}
	\centering
	\begin{tabular}{m{1cm} m{3cm} m{2.5cm} m{1cm}}
		\hline
		\hline
		$T$ & P:$\max \{ \tau_\mathrm{w} \}$ &  V:$\max \{ \tau_\mathrm{w} \}$ & $t_\mathrm{ob}$\\
		\hline
		0.32 & 1135.76(1142.34) & 0.75 &1142.34\\
		\hline
		0.28 & 1220.42(1224.47) & 3.54 &1224.47\\
		\hline
		0.22 & 1204.43(1204.75) & 83.87 &1204.75\\
		\hline
		\hline
	\end{tabular} \label{tb:de}
\end{table}
\begin{figure}[!th]
\includegraphics[width=\columnwidth]{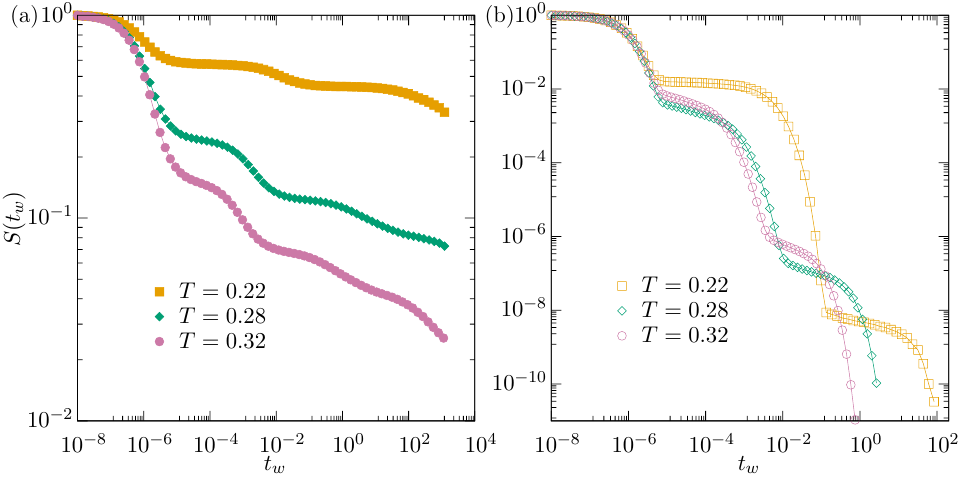}
\caption{The survival function $S(t_\mathrm{w})$ for particles (a) and vacancies (b). The maximum waiting time
$t_\mathrm{w}$ of vacancies is less than $10\%$ of the total measurement time. The solid lines are guide to the eyes.}\label{fig:sf}
\end{figure}

Finally, we argue that the motion of vacancies becomes diffusive at the long-time limit. However, direct demonstration by the
slope of MSD at the long time is not feasible given the computational cost at low $T$. For this purpose, let us study the empirical survival function $S(t_\mathrm{w})$ given by,
\begin{equation}
S(t_\mathrm{w})=\overline{\left \langle  \frac{1}{K_i}\sum_{j=1}^{K_i} \Theta (\tau_{\mathrm{w},j}^{i}-t_\mathrm{w}) \right \rangle}, 
\end{equation}
where $\tau_{\mathrm{w},j}^{i}$ labels the $j-$th waiting time of the $i$-th P/V, $K_i$ is the total number of the
collected waiting times for the $i$-th P/V. $\Theta(x)$ is the Heaviside function. We have collected $\tau_{\mathrm{w},j}^{i}$ in a
hopping-event-driven fashion.  All the hopping events are recorded in this series of simulations and the waiting times are then extracted.  Hence $
S(t_\mathrm{w})$ measures the probability that the P/V exhibits no hopping motion till
$t_\mathrm{w}$ and its derivative with respect to $t$ yields the waiting time distribution. Table \ref{tb:de} lists the maximum
observation time window and the maximum waiting times. The maximum $\tau_{\mathrm{w},j}^{i}$ of vacancies is less than
10\% of the total measurement time. In contrast, some particles remain inactive in the whole process. \fig{fig:sf}(a) and (b)
plot $S(t_\mathrm{w})$ for particles and vacancies respectively. For best clarity, a log-log coordinate is used. No
obvious power-law tail can be identified.  The
two-stage slowdown in MSD and $q(t)$ can also be observed in $S(t_\mathrm{w})$, and it lasts about one decade at $T=0.32$ and
grows to more than two decades at $T=0.22$. Since we have adopted the continuous time algorithm~\cite{Bortz_1975} to
study the hopping dynamics, it is convenient to consider the diffusion in the generalized continuous time 
random walk (CTRW) framework~\cite{Bouchaud_1990}. In order to apply the central limit
theorem(CLT)~\cite{Salgado_Garc_a_2013,Chazottes_2015}, it is crucial to ensure
the fast decay of the correlation in the system and a narrow distribution of waiting times between consecutive hops.
Here, the waiting times of vacancies
are finite. It is also clear that $C(k\tau)$ decays to zero for $k \le 15$. Satisfaction of these conditions strongly
suggests vacancies' normal diffusion at the long-time limit.  As for particles, a slowly-decaying tail cannot be ruled out
by our numerical evidence, and its fate at the long-time limit remains unknown.

A two-stage slowdown has been demonstrated in both the MSD and $q(t)$. If we naively take the rise from the first plateau as the secondary $\beta$-relaxation~\cite{Goldstein_1969,Johari_1970a,Johari_1970b,Shiraishi_2023} and the rise from the second plateau as the primary $\alpha$-relaxation~\cite{Shiraishi_2023}, a hierarchy of spatial scale can be mapped to the motion of vacancies. The secondary process is characterized by frequent transitions of vacancies between few localized regions and the primary relaxation involves isolated long-waiting-time hops to escape the larger trap of Type II. Though a consensus is not yet reached regarding the the underlying
mechanism for $\beta$ and $\alpha$
relaxation~\cite{Gotze_1992,Ngai_1998,Ngai_2004,Stevenson_2009,Johari_2019,Gao_2025}, the view of a separation of length scale is generally shared. The
hierarchical relaxation exhibited by our model appears naturally
consistent with the rugged potential energy landscape (PEL) of
Stillinger~\cite{Stillinger_1995}. It is tempting to associate the
hopping motion at different length scale to the intrabasin and
intercrater diffusion on the PEL, though more quantitative work is needed.

Sharing the same merit of simplicity of  lattice-gas models~\cite{Fredrickson_1985,Nicodemi_1997a,Nicodemi_1997b,Arenzon_2000,Fierro_2000,de_Candia_2001,Denny_2003,Ritort_2003,Bertin_2005,Tondl_2014,Zhang_2017,Nishikawa_2024} for glassy systems, our model is a natural derivative of the Edwards-Anderson model~\cite{Edwards_1975} in 2D. Closely related is the frustrated lattice gas model~\cite{Nicodemi_1997a} in finite dimension and $p$-spin models~\cite{Crisanti_2011} in the infinite dimension, which all demonstrate glassy dynamics and originate in spin-glass models. The emergence of glassy dynamics induced by the elements of frustration and crowdedness corroborates the idea of existence of deeper connection between spin glasses and other general glassy systems~\cite{Kirkpatrick_1989, Parisi_2010}. Though the disorder of structural glasses is believed to be geometrically self-generated, quenched disorder in this model leads to similar diffusion and relaxation phenomena. Again this may inspire more quantitative studies focusing on of their similarities with regard to the energy landscape in finite dimension.

In summary, we have presented a lattice gas model with spatially-quenched interactions and particle-quenched spins.  In particular, we observe a two-stage slowdown for the diffusion and structural relaxation. We ascribe this slowdown to localization and anti-correlation at different spatial and temporal scale. Further study is needed concerning the fate of particle diffusion at the long-time limit.

\acknowledgments
This work was supported by National Natural Science Foundation of China Grant No. 12405042. 

\bibliography{two_sd.bib} 

\end{document}